# Stretchable liquid crystal blue phase gels

F. Castles[1,2]*, S. M. Morris[1,3], J. M. C. Hung[1], M. M. Qasim[1], A. D. Wright[1], S. Nosheen[1], S. S. Choi[1], B. I. Outram[3], S. J. Elston[3], C. Burgess[4], L. Hill[4], T. D. Wilkinson[1] and H. J. Coles[1]

[1] Centre of Molecular Materials for Photonics and Electronics, Department of Engineering, University of Cambridge, 9 JJ Thomson Avenue, Cambridge CB3 0FA, UK.
[2] Department of Materials, University of Oxford, Parks Road, Oxford OX1 3PH, UK.
[3] Department of Engineering Science, University of Oxford, Parks Road, Oxford OX1 3PJ, UK.
[4] Defence Science & Technology Laboratory, Porton Down, Salisbury SP4 0JQ, UK.

**Liquid crystalline polymers are materials of considerable scientific interest and technological value to society[1-3]. An important subset of such materials exhibit rubber-like elasticity; these can combine the remarkable optical properties of liquid crystals with the favourable mechanical properties of rubber and, further, exhibit behaviour not seen in either type of material independently[2]. Many of their properties depend crucially on the particular mesophase employed. Stretchable liquid crystalline polymers have previously been demonstrated in the nematic, chiral nematic, and smectic mesophases[2,4]. Here were report the fabrication of a stretchable gel of blue phase I, which forms a self-assembled, three-dimensional photonic crystal that may have its optical properties manipulated by an applied strain and, further, remains electro-optically switchable under a moderate applied voltage. We find that, unlike its undistorted counterpart, a mechanically deformed blue phase exhibits a Pockels electro-optic effect, which sets out new theoretical challenges and new possibilities for low-voltage electro-optic devices.**

Liquid crystalline polymers are important in diverse areas of science and technology; for example, the liquid crystalline properties of DNA were relevant in the experiments which led to the discovery of its structure[5], and the majority of today's high-performance display devices now contain a liquid crystalline polymer network[3]. The subset of liquid crystalline polymers that exhibit rubber-like elasticity are particularly interesting in some respects as they display rich new physics, such as soft elasticity[2,6], and promise novel technologies, such as artificial muscles[7]. Given that stretchable nematics, chiral nematics, and smectics each have unique properties and potential device applications[2], we believed it would be fruitful to investigate whether stretchable blue phases are also possible. In particular, we were initially motivated by the fact that the exotic three-dimensional (3D) structures of blue phases I and II suggest new properties and device capabilities in the context of stretchable 3D photonic crystals.

Blue phase polymers have previously been investigated in non-crosslinked systems[8-10], including concentrated DNA solutions[11], and in crosslinked networks[12]. In recent years, blue phase gels—often referred to as 'polymer stabilised' blue phases[13]—have received considerable attention because they enable the wide-temperature operation of electro-optic devices that exploit the blue phase's remarkably large Kerr constant, which is typically $10^2$ to $10^4$ times larger than that of nitrobenzene[14,15]. Major display manufacturers have produced prototype televisions based on the effect[16]. However, dense networks of the type reported in ref. 12 and gels of the type reported in ref. 13 do not withstand any substantial mechanical deformation: the former are glassy while the latter lack sufficient structural integrity. Polysiloxane blue phases of the type reported in ref. 10 may provide a viable route to stretchable blue phases, if suitably crosslinked, though this would remain to be demonstrated—in any case, only multi-domain samples appear to be possible in such systems at present, whereas many of the interesting properties of blue phases are limited to mono-domain samples. Further, in polysiloxane elastomers the characteristic ability of low molar mass liquid crystals to have their optical properties readily switched by an electric field is typically supressed, particularly if the sample is mechanically constrained[2].

To obtain a blue phase polymer that is substantially stretchable and exhibits large mono-domains, yet remains readily electro-optically switchable in an applied electric field, we formed gels via the in-situ photopolymerisation of a mixture that included mono- and di-acrylate reactive mesogens (Methods). The in-situ photopolymerisation of reactive mesogens, pioneered by Broer et al.[17], has previously been employed to stabilise a wide variety of liquid crystals—including the blue phases—and to increase their functionality (e.g., refs. 3,12-20). The technique has been reported as a route to creating mono-domain chiral nematic elastomers (see, e.g., ref. 3, p. 369). Our premixtures contained ≈ 30 wt% reactive mesogens in a non-reactive mesogenic host mixture; it is the intermediate 'extent of polymerisation', determined primarily by the amount of reactive mesogens used and the degree of exposure to ultraviolet light, which enables our blue phase materials to be stretched, unlike dense blue phase networks[12] (≈ 100 wt% reactive mesogens) or polymer stabilised blue phases[13] (≈ 5 wt% reactive monomers). Our approach was facilitated by recent advances in blue phase materials with naturally wide temperature ranges[21] which were robust under the required processing procedure (Methods). However, such materials are not, in principle, essential. After photopolymerisation, the materials appeared macroscopically as solids—that is, they retained their shape—as shown in Fig. 1. The structure was identified conclusively as that of blue phase I using a combination of standard techniques—polarising optical microscopy, spectral analysis, and Kossel diffraction analysis—and was stable over a wide temperature range, including room temperature.

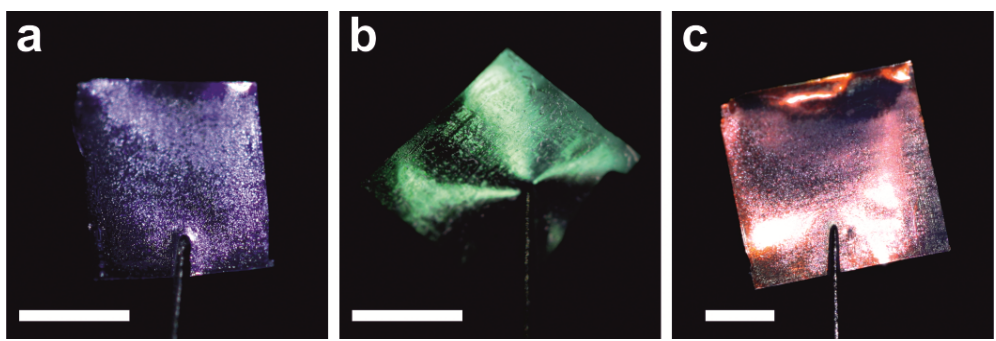

**Figure 1 | Free-standing films of a blue phase gel**. The films appear coloured due to selective reflection of visible light (scale bars, 3 mm). The three samples, **a**, **b**, and **c**, have the periodicity of their structure chosen such that they reflect blue, green, and red light respectively.

Figure 2a-c and Supplementary Videos 1 & 2 show the blue phase gels being stretched. Here, a ≈ 20 μm–thick layer was placed across a variable-width aperture and observed on a polarising optical microscope. For demonstration purposes, in this case, characteristic platelet textures were formed by choosing a suitable cooling rate. The film adhered to the aperture edges and was stretched as the aperture was widened. So far we have succeeded in stretching the films in this way by up to ≈ 1.5–2 times their original length. This is currently considerably less than the maximum strains reported for some elastomer systems in other mesophases[2]. Nevertheless, it is sufficient to prove our concept and, as we will see, to produce strain-induced colour changes over almost the entire region of visible wavelengths. The distortion was carried out at room temperature and is reversible, as seen clearly in Supplementary Video S2. The behaviour is also repeatable over many stretch-relaxation cycles (~ 10 stretch-relaxation cycles were tested). In Supplementary Video S1 the film is stretched until it breaks, with the subsequent recoil demonstrating rubber-like elasticity.

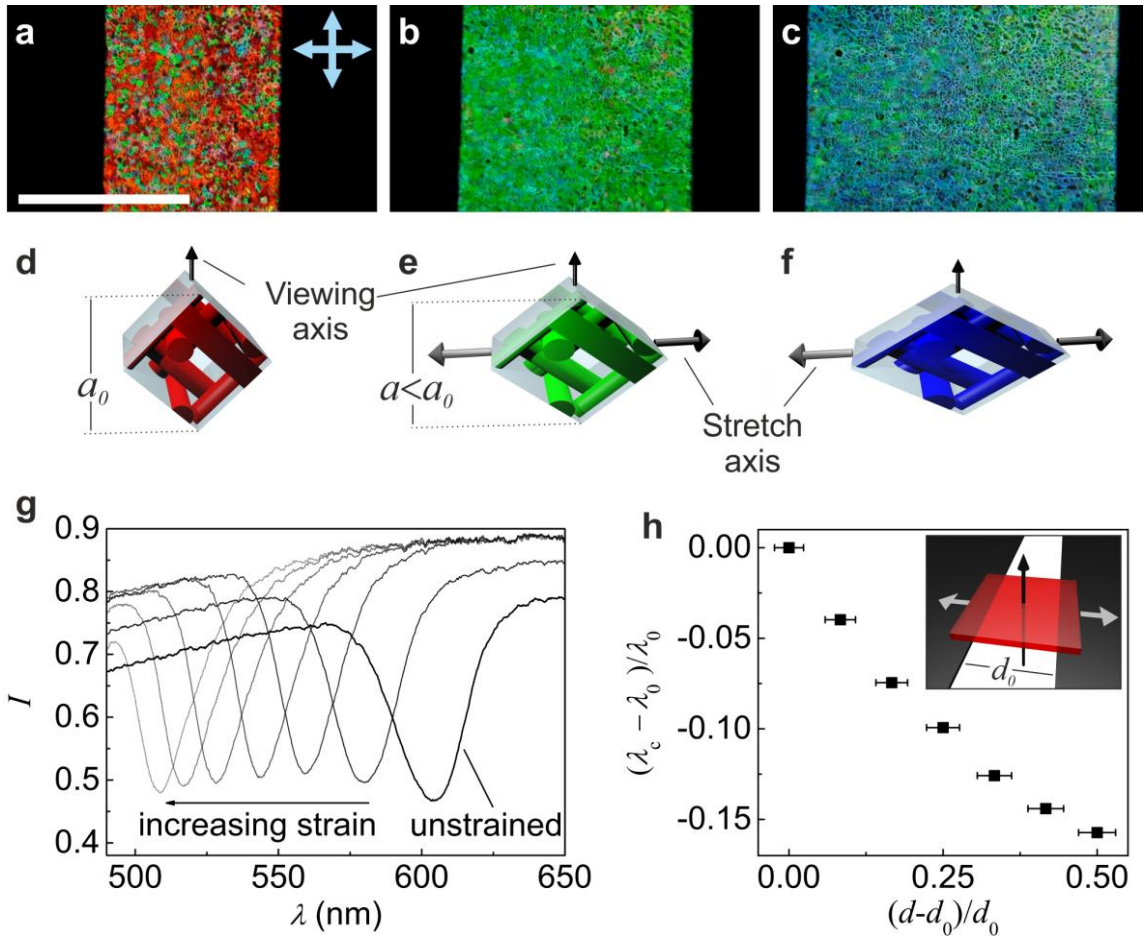

**Figure 2 | Colour changes in a stretchable blue phase gel**. **a**, **b** & **c**, Polarising optical microscope images of a 20 μm thick layer placed across a variable-width aperture. **a**, The un-stretched state displays a characteristic blue phase texture. Pale blue arrows denote the orientation of the crossed polarisers (scale bar, 1 mm). **b** & **c**, As the aperture is widened the sample is stretched and a colour change induced. **d**, **e** & **f**, Schematic diagrams of a typical deformation of the blue phase I unit cell, viewed along the [011] direction and stretched along an axis perpendicular to this. **g**, Spectral analysis of a sample with uniformly aligned [011] platelets showing the relative transmitted intensity $I$ as a function of the vacuum wavelength of transmitted light $\lambda$. Curves with reflection bands from right to left correspond to aperture widths in this experiment of 0.60 mm (unstrained), 0.65 mm, 0.70 mm, 0.75 mm, 0.80 mm, 0.85 mm, and 0.90 mm respectively. **h**, The wavelength of maximum selective reflection $\lambda_c$, defined as the minimum of $I(\lambda)$, is plotted as a function of the aperture width $d$, normalised by the values in the initial, unstrained, state $\lambda_0$ and $d_0$. The random uncertainty in a measurement of $d$ was estimated from the graduation on the screw gauge to be ± 0.01 mm, and the random uncertainty in the measurement of $\lambda_c$ was negligible in relation to this. Inset: schematic diagram showing the geometry of the experimental setup.

Upon stretching, a colour change is readily apparent in the polymer films. Blue phases may be considered as photonic crystals with stop bands for certain polarisations and frequencies of visible light; the observed mechano-chromic behaviour is consistent with the imposed lateral stretch inducing a reduction of the layer thickness and hence a reduction of effective photonic crystal lattice periodicity along the viewing direction, due to approximate volume conservation of the material. This is illustrated schematically in Fig. 2d-f. The effect is analogous in many respects to the one-dimensional case of chiral nematic elastomers[22]. To quantify the colour change we recorded the transmission spectra as the aperture was widened, as shown in Fig. 2h. Here, rather than forming a characteristic platelet texture, the blue phase was formed as per Fig. 1 such that it displayed only [011] platelets and hence exhibited a single, narrow, region of selective reflection at visible frequencies. The blue-shift of the selective reflection region observed in Fig. 2g&h is qualitatively concomitant with the colour change observed in Fig. 2a-c, although the particular sample used for Fig. 2g&h did not withstand the same extent of deformation before breaking. We also note that, although it is not readily apparent in Fig. 2a-c because the stretch axis was aligned with a polariser axis, the samples exhibited strain-induced birefringence (Supplementary Information).

We initially sought to investigate the Kerr effect in stretchable blue phases. However during the course of our investigations we made an unexpected discovery of potentially much greater interest: a qualitatively new electro-optic response in distorted blue phases that is not observed in their undistorted counterparts. The lowest order electro-optic effect observed previously in the blue phases is the quadratic (Kerr) effect[14,23]. Kitzerow has theoretically investigated polar electro-optic effects beyond the Kerr effect, identifying flexoelectricity as a potential mechanism, but he concludes that "the flexoelectric effect in the blue phases is expected to have little effect on the field-induced birefringence, if any"[24]. Indeed, the arguments of classical crystal optics forbid the existence of a linear electro-optic (Pockels) effect in the blue phases on account of their macroscopic symmetries[23,25]; blue phases I and II belong to the *O(432)* crystal class and blue phase III is amorphous[24]. While it should be borne in mind that this argument applies in the limit that the wavelength of light in the medium is much larger than the periodicity of the structure—which is true in conventional crystal optics but not necessarily true for blue phases observed using visible light—it is nevertheless the case that the electro-optics of the blue phases have been extensively investigated and no Pockels effect has previously been reported. To motivate the existence of qualitatively new electro-optics in distorted blue phases we suggest that, since *O(432)* is the only non-centrosymmetric crystal class for which every component of the linear electro-optic tensor is simultaneously zero, typical distortions of blue phase I or blue phase II can break the symmetry which 'forbids' a classical Pockels effect. Therefore, a Pockels effect is possible in distorted blue phase I or II at all wavelengths.

To demonstrate the effect experimentally we report here results for a distorted gel of blue phase I fabricated using a templating technique (Methods). This sample is the same, in principle, to the stretchable materials described above; that is, it is a blue phase gel formed from a premixture that contained ≈ 30 wt% reactive mesogens. It was used because it exhibited a small amount of natural distortion—perhaps on account of swelling/de-swelling during solvent exchange or through interaction with the surface of the cell—which was apparent through the observation of a small intrinsic birefringence ~$10^{-3}$ (Supplementary Information). Thus, by employing this sample, it was not necessary to remove the material from the cell, stretch it, and then reattach electrodes. Further, the sample was very robust and provided measurements that were accurately repeatable over a period of many months. The cell was placed between crossed polarisers and the transmitted light was measured using a photodiode as a function of the applied voltage (Methods). The light source was a red fibre-coupled high-power light emitting diode with peak intensity at 640 nm and a full width at half maximum of 19 nm. This wavelength was chosen to avoid potential complications arising from diffraction at the region of selective reflection, which was determined by spectral analysis to be centred at 540 nm with a full width at half maximum of 21 nm. With the frequency of the sinusoidal driving voltage set at $f = 113$ Hz, a lock-in amplifier was used to measure separately the 1$^{st}$ harmonic ($f$) and 2$^{nd}$ harmonic ($2f$) of the photodiode response (Methods). The results are shown in Fig. 3, and a clear Pockels effect is apparent in Fig. 3a.

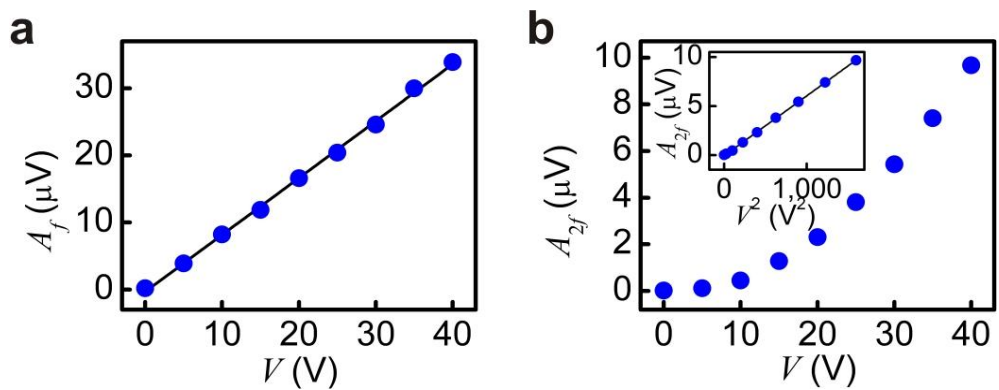

**Figure 3 | Electro-optics of a distorted blue phase gel**. Harmonic analysis of the rms amplitude of the photodiode response as a function of the rms amplitude of the applied voltage $V$. **a**, The first harmonic $A_f$ at the same frequency as the applied voltage is linear in $V$, demonstrating the existence of a linear electro-optic (Pockels) effect. **b**, The second harmonic $A_{2f}$ at twice the frequency of the applied voltage is quadratic in $V$, consistent with a quadratic electro-optic (Kerr) effect. Each data point represents the mean of ten repeated measurements. The standard uncertainties on the means were found to be 1 μV for all $A_f$ data points, and 0.1-0.2 μV for all $A_{2f}$ data points, which is within the thickness of the data symbols.

As a control experiment—i.e., to confirm that the observed Pockels effect is an interesting consequence of the distorted blue phase structure and not a spurious result due, say, to ion flow—we carried out the same electro-optic experiment on a non-polymer blue phase sample (Methods). If the sample was given sufficient time to relax to an undistorted state, no significant birefringence or Pockels effect was observed above the experimental uncertainty in the measurement. Further, it was possible to induce a significant Pockels effect in the same sample by distorting the blue phase structure—achieved simply by prodding the cell with a pair of tweezers. After inducing such a distortion, the size of the Pockels effect—quantified by the gradient of $A_f(V)$—would reduce on a time scale of some tens of minutes, which is consistent with relaxation to an undistorted state. This provides very strong evidence that the Pockels effect reported here is a distortion-induced effect.

The experimental realisation of stretchable blue phase polymers poses new scientific challenges; the theory of nematic, chiral nematic, and smectic elastomers and gels has been actively investigated for some decades[2], but the theory of stretchable blue phase polymers is yet to be explored. Certainly, the experimental discovery of a Pockels effect in distorted blue phases sets out one specific new experimental and theoretical challenge in particular; while we have argued that the effect is permissible according to the symmetry arguments of classical crystal optics, the microscopic origin of the effect in terms of given director or Q-tensor fields is now an open question. A detailed analysis of the problem should be amenable to modern theoretical methods[26-28] but is beyond the scope of this Letter. The concept that distortion-induced symmetry-breaking permits the observation of otherwise symmetry-forbidden phenomena in the blue phases may be extended and allow us to *predict* the existence of piezoelectricity in blue phase I or II if some cross-linking is carried out in a distorted state.

The new electro-optics of distorted blue phases, reported here, may be of interest from a technological perspective. As noted above, the unusually large Kerr constant of cubic blue phases has made them a viable material for electro-optic devices such as displays[13-16]. What is particularly interesting is that, since the Pockels effect is linear in the applied electric field rather than quadratic, it necessarily dominates over the Kerr effect at low voltages, as is apparent in Fig. 3. Thus stretchable blue phases may be of interest for future low-voltage electro-optic devices.

**Methods**

**Materials and processing procedure.** Premixtures for the stretchable blue phases shown in Figs. 1 & 2, Fig. S2 (Supplementary Information), and Supplementary Videos S1 & S2 were typically composed of 4 wt% chiral additive BDH1281 (Merck), 30 wt% reactive-mesogen/photoinitiator mixture UCL-011-K1 (DIC Corp.), 16.5 wt% $B_5$, 16.5 wt% $B_7$, 16.5 wt% $B_9$, and 16.5 wt% $B_{11}$, where the chemical structure of the B-series of bimesogens, synthesized in-house, is shown in Fig. S3 (Supplementary Information). The periodicity of the lattice, and hence the observed colour (e.g., Fig. 1), was tuned by varying the amount of chiral additive in the range 3.3–4.4 wt%. Mixtures were capillary-filled between two sheets of glass, forming a layer 20 μm thick. The samples used for Fig. 1 and Fig. 2g&h used commercial cells with planar alignment layers (Instec) while all other samples used cells made in-house with no alignment layers. The blue phase was formed by cooling slowly from the isotropic phase and was polymerised under ultraviolet radiation, typically of intensity 50 W/m$^2$. A two-step process was employed to form stretchable samples: First, the blue phase was exposed for 7 s at an elevated temperature. Second, it was exposed for 13 s at room temperature. After removal of one glass layer, the film was separated from the other glass layer using a razor blade wetted with propan-2-ol. All reported images and data concerning the stretchable samples were obtained at room temperature.

**Electro-optic characterisation.** The material and preparation procedure for the sample used in Fig. 3 and Fig. S2 (Supplementary Information) are the same as those reported in ref. 19 (Premixture 1, refilled with 5CB). To study the electro-optic properties the cells were investigated, in the first instance, on a polarising optical microscope in the Cambridge laboratories. Characterisation was carried out on single platelets which filled the entire viewing area of the microscope with a 40× objective lens. The experiment focused on platelets that were oriented with the [011] crystallographic direction aligned normal to the cell surface and parallel to the viewing direction, as confirmed by comparing the resulting Kossel diffraction diagrams against the known literature (e.g., ref. 29). A signal generator and a high-voltage amplifier were used to apply a voltage across transparent indium-tin-oxide electrodes on the inner surfaces of the cell, creating an electric field parallel to the direction of observation. The transmitted light was detected using a photodetector (PDA36A-EC, Thorlabs) mounted on the microscope. Accurate crossing of the polariser and analyser was confirmed. In this configuration the dominant contribution of the Kerr effect—that is, induced birefringence with symmetry axis along the field direction—does not affect the transmitted intensity, allowing any contribution of the Pockels effect to be revealed more readily. However, a small variation in the transmitted intensity due to the Kerr effect may still be expected since an electric field applied along the [011] direction of an $O(432)$ material will, in general, induce biaxiality[23-25]. We believe that this biaxial Kerr effect is the dominant contribution in Fig 3b. However, we have not experimentally ruled out or quantified the contribution from the effect discussed in ref. 30, whereby electrostatic attraction of the electrodes reduces the device thickness—this could possibly induce an elasto-optic effect in the cubic blue phases that may be observable in our experiment. Yet, on theoretical grounds one may expect this effect to be negligible, particularly at the low electric field values employed here (less than approximately 2 V/μm). Note, in addition, that such variations in the cell thickness cannot contribute to the linear electro-optic effect in Fig. 3a, since the electrostatic attraction of the electrodes is independent of the polarity of the electric field (and, in any case, the control experiment, described above, rules out potential spurious contributions such as this). Applied voltages were kept below 40 Vrms, over a cell gap of ≈ 20 μm, ensuring dielectric breakdown did not occur. At such voltages the total field-induced change in transmitted light was perceptible to the naked eye when viewed down the microscope, but barely so; the relative changes in transmitted intensity were ~ 0.1% when normalised to the intensity of light transmitted through parallel polarisers with no sample present (the photodiode output voltage for parallel polarisers was 380 mV for the experiment reported in Fig. 3). A lock-in amplifier ensured accurate measurements even at low applied voltages $V$. A frequency of 113 Hz was chosen because it is of the same order of magnitude as consumer electronics, yet avoids potential mains interference. To confirm that the result was robust, the experiment was repeated under a number of different conditions: First, it was repeated on the polarising optical microscope setup using light of different wavelengths—in total, eight fibre-coupled high-power light emitting diodes (Thorlabs) with nominal wavelengths of 420, 455, 530, 565, 590, 625, 660, and 780 nm were tested. Second, the experiment was repeated in the Oxford laboratories using a modified setup; the sample was mounted between crossed polarisers on an optical bench and a 633 nm helium-neon laser was used as the light source. A linear electro-optic effect was confirmed in all cases. We found that the value of $A_f$ varied as the sample was rotated about an axis defined by the viewing direction (Supplementary Information); Fig. 3 reports data for the orientation which maximised $A_f$. All electro-optic data concerning this sample were obtained at room temperature.

**Acknowledgments**

This work was funded by the Engineering and Physical Sciences Research Council UK under the COSMOS project (grants EP/D04894X/1 and EP/H046658/1), and by the Defence Science & Technology Laboratory UK. We thank A. Lorenz, J. Montelongo, D. Gardiner, K. Knowles, and L. Tian for useful discussions, and H. Hasebe (DIC Corp., Japan) for supplying the material UCL-011-K1. S.M.M. acknowledges The Royal Society for financial support.

# Supplementary information

Birefringence of distorted blue phases

Supplementary Fig. S1a, b, & c show polarising optical microscope images of a stretchable blue phase sample in the undistorted state as it is rotated with respect to the orientation of the crossed polarisers. Essentially no variation in the transmitted intensity is observed, as expected. Supplementary Fig. S1d, e, & f show the same sample in a strained state: in this case the transmitted intensity varies as the sample is rotated, demonstrating that the distorted state is birefringent.

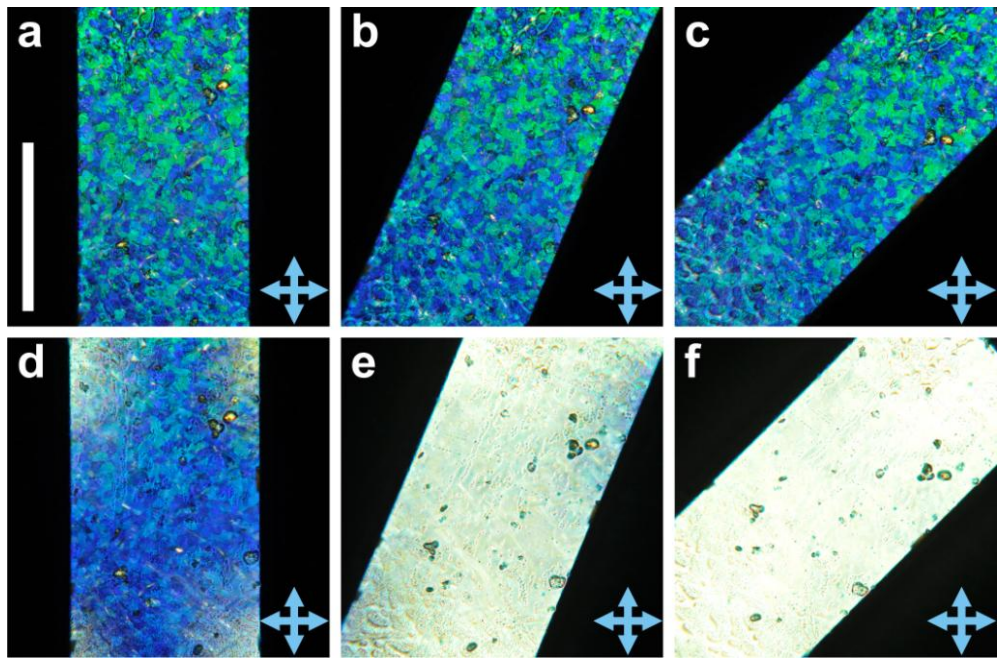

**Supplementary Figure S1 | Strain-induced birefringence in a stretchable blue phase.**

Intrinsic birefringence and further electro-optics of the distorted blue phase gel

Here we provide evidence that the templated blue phase that was used to demonstrate a Pockels effect in Fig. 3 is a distorted blue phase structure and determine its birefringence. The birefringence of a [011] platelet of the templated blue phase sample is characterised in Supplementary Fig. S2 (upper plot); the normalised transmitted intensity between crossed polarisers $I$ is plotted as a function of the angle of rotation $\theta$, showing that $I$ is approximately proportional to $\sin^2(2\theta)$, which is consistent with a birefringent sheet. Rotation was, again, about an axis defined by the viewing direction. The light source used in Supplementary Fig. S2 is the same 640 nm light emitting diode referred to in the main text. The light leakage in the dark state, i.e., the nonzero values of the minima of $I(\theta)$, is consistent with optical activity; the leakage was substantially reduced if the analyser was slightly uncrossed, and, further, it was found that the sense of rotation required to achieve such a reduction was opposite for wavelengths of light on opposite sides of the region of selective reflection. By equating the modulation in the relative transmitted intensity ($\sim 10^{-2}$) to $\sin^2(\pi d \Delta n/\lambda)$, using $d$ = 20 μm and $\lambda$ = 640 nm, gives $\Delta n \sim 10^{-3}$.

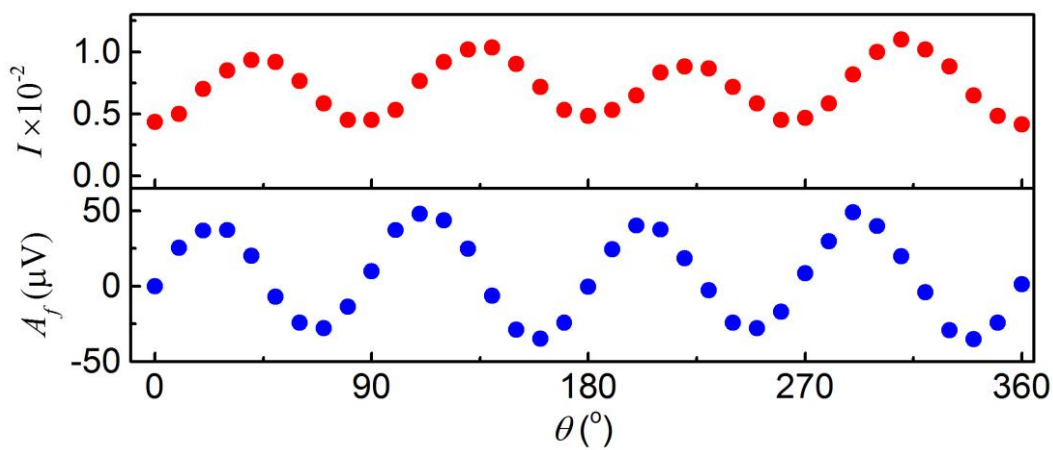

**Supplementary Figure S2 | Birefringence and electro-optics of a distorted blue phase gel. a**, The relative transmitted intensity $I$ (normalised to the intensity through parallel polarisers with no sample present) as a function of the angle of rotation of the sample $\theta$. **b**, The Pockels effect, quantified by the photodiode response at the frequency of the applied voltage, $A_f$, as a function of the angle of rotation.

We note that care must be taken in inferring distortion from a *small* observed birefringence because even undistorted blue phases may exhibit linear birefringence near the region of selective reflection[1-4] (the common argument, borrowed from classical crystal optics, that cubic symmetry implies optical isotropy, and vice-versa, is not necessarily valid in this region[5]). The signature of such intrinsic birefringence is predicted to be a strong frequency dependence and a change in its sign at the region of selective reflection[1,3]. We observed the birefringence in our sample at a range of wavelengths of incident light and, away from the region of selective reflection, where the results of Fig. 3 and Supplementary Fig. S2 were obtained, we observed only a weak frequency dependence, consistent with the frequency dispersion of $\Delta n$ for 5CB. Further, away from the region of selective reflection we found that $\Delta n$ was of the same sign on each side. Therefore we conclude that the birefringence observed in our templated sample at the measurement wavelength of 640 nm is not due to the intrinsic birefringence discussed in refs. 1-4, but due instead to distortion of the sample. Birefringence of blue phase samples due to surface distortion is known and has been discussed previously, for example, in ref. 6.

Rotational dependence of the Pockels effect

It was found that the strength of the Pockels effect, characterised by $A_f$, varied as the sample was rotated about an axis defined by the viewing direction, as shown in Supplementary Fig. S2 (lower plot). It appears that, approximately, $A_f$ is proportional to $dI/d\theta$, which is consistent with an electro-optic effect arising from a small in-plane rotation of the optic axis.

Non-polymer sample & chemical structures of bimesogens

The mixture for the electro-optic experiment on a non-polymer sample was composed of 3.6 wt% chiral additive BDH1281 (Merck), 13 wt% $A_7$, 34 wt% $A_9$, 13 wt% $A_{11}$, 5 wt% $B_5$, 5 wt% $B_7$, 5 wt% $B_9$, 5 wt% $B_{11}$, 6 wt% $C_9$, and 10 % viscosity diluter (Merck), where the A, B, and C series bimesogens were synthesised in house and are shown in Supplementary Fig. S3. Electro-optic experiments were again carried out on [011] platelets of blue phase I which were large enough to fill the entire visible area of the microscope with a 40× objective lens. Experiments on this sample were carried out for blue phase I at 55.5 °C.

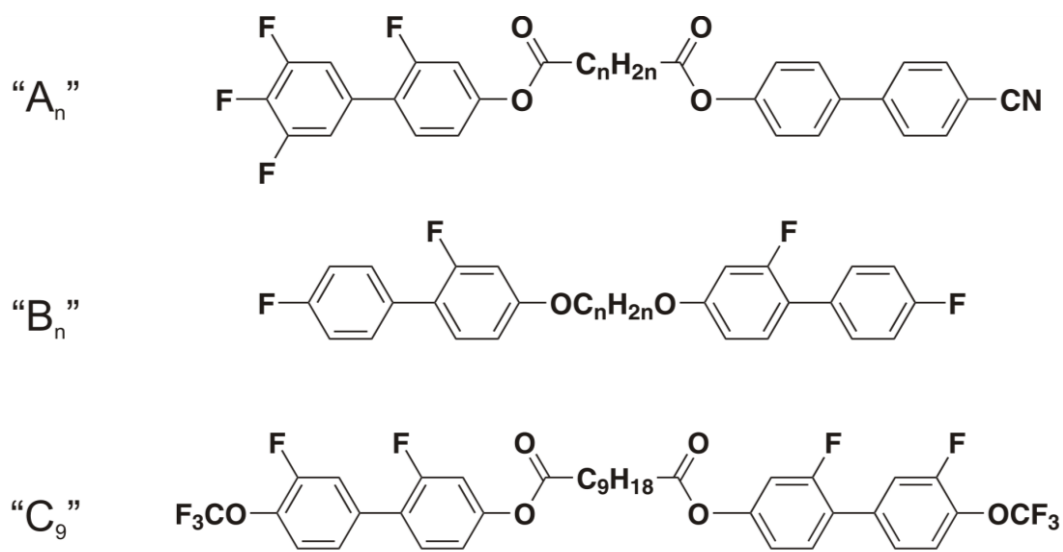

**Supplementary Figure S3 | Chemical structures of the bimesogens used in the study.**